\documentclass[sigconf,9pt,usenames,dvipsnames,table]{acmart}

\AtBeginDocument{%
  \providecommand\BibTeX{{%
    \normalfont B\kern-0.5em{\scshape i\kern-0.25em b}\kern-0.8em\TeX}}}

\setcopyright{rightsretained}
\copyrightyear{2024}
\acmYear{2024}
\acmDOI{10.48550/arXiv.2411.16340}
\acmConference[LOCO '24]{LOCO '24: 1st International Workshop on Low Carbon
Computing}{December 03, 2024}{Glasgow, UK}
\acmBooktitle{LOCO '24: 1st International Workshop on Low Carbon 
Computing, December 03, 2024}
\acmISBN{}

\usepackage{balance}
\usepackage{verbatim}
\usepackage[textsize=footnotesize, textwidth=15mm]{todonotes}
\presetkeys{todonotes}{fancyline}{}
\usepackage[english=american,babel=false]{csquotes}
\usepackage{nth}

\usepackage{acronym}
\acrodef{EU}{European Union}

\definecolor{bw1}{RGB}{0,114,178}
\definecolor{bw2}{RGB}{230,159,0}
\definecolor{bw3}{RGB}{86,180,233}
\definecolor{bw4}{RGB}{0,158,115}
\definecolor{bw5}{RGB}{240,228,66}

\makeatletter
\newcommand*\bigcdot{\mathpalette\bigcdot@{2}}
\newcommand*\bigcdot@[2]{\mathbin{\vcenter{\hbox{\scalebox{#2}{$\m@th#1\bullet$}}}}}
\makeatother

\begin{document}

\title[Exploring Privacy and Security
  as Drivers for Environmental
  Sustainability in Cloud-Based
  Office Solutions]{Exploring Privacy and Security
  as Drivers for Environmental
  Sustainability in Cloud-Based
  Office Solutions \\ (Extended Abstract)}

\author{Jason Kayembe}
\email{jason.kayembe@ulb.be}
\affiliation{%
  \institution{Universit\'e Libre de Bruxelles}
  \city{Brussels}
  \country{Belgium}
  \postcode{1050}
}

\author{Iness Ben Guirat}
\email{iness.ben.guirat@ulb.be}
\affiliation{%
  \institution{Universit\'e Libre de Bruxelles}
  \city{Brussels}
  \country{Belgium}
  \postcode{1050}
}

\author{Jan Tobias M\"uhlberg}
\email{jan.tobias.muehlberg@ulb.be}
\orcid{0000-0001-5035-0576}
\affiliation{%
  \institution{Universit\'e Libre de Bruxelles}
  \city{Brussels}
  \country{Belgium}
  \postcode{1050}
}

\begin{abstract}
  This paper explores the intersection of privacy, cybersecurity, and
environmental impacts, specifically energy consumption and carbon
emissions, in cloud-based office solutions. We hypothesise that solutions
that emphasise privacy and security are typically \enquote{greener} than
solutions that are financed through data collection and advertising. To
test our hypothesis, we first investigate how the underlying architectures
and business models of these services, e.g., monetisation through
(personalised) advertising, contribute to the services' environmental
impact. We then explore commonly used methodologies and identify tools that
facilitate environmental assessments of software systems. By combining
these tools, we develop an approach to systematically assess the
environmental footprint of the user-side of online services, which we apply
to investigate and compare the influence of service design and ad-blocking
technology on the emissions of common web-mail services. Our measurements
of a limited selection of such services does not yet conclusively support
or falsify our hypothesis regarding primary impacts. However, we are
already able to identify the greener web-mail services on the user-side and continue the
investigation towards conclusive assessment strategies for online office
solutions.

\end{abstract}


\maketitle
\renewcommand{\shortauthors}{Kayembe et al.}


\section{Introduction}

Following growing concerns over the environmental impacts of human
activities across all sectors, this paper investigates the intersection of
privacy, security, and the environmental consequences of ICT services, with
a particular focus on web-based email services as an example of commonly
used
office solutions. We explore how to define and evaluate
these services, examine the architecture and business models they rely on,
and envision a framework that can measure the energy consumption associated
with specific email services. Our research hypothesis is that \emph{online
  services that emphasise privacy and security are typically
  \enquote{greener,}} and
we take web mailers as a case study to test this hypothesis.

Sending an email might seem like a trivial task; however the scale at which
large email services operate makes their environmental impact significant.
For instance,
according to Statista, a web platform for markets statistics, an estimated
333 billion emails were sent and received daily worldwide in 2022~\cite{statistaemails2024}.
Perhaps even more important, beyond the data volume, is that the infrastructure supporting email services requires many components of the internet network to remain operational which, in return consumes energy and requires maintenance continuously.
This is illustrated in~\cite{parssinen_environmental_2018} where P\"arssinen et al. take inventory of common internet network devices
to derive the overall environmental impact of the online-advertising
industry.
An important factor influencing the environmental impacts of the different
services is the business model behind the services. We identify six key
models: 1. subscription-based, 2. transaction-based, 3.  freemium, 4.
donation-based, 5. advertisement-based, and 6. monetisation through data
collection and reselling. Specifically, services that rely on advertising and
data collection often deploy additional technologies that go beyond
email
functionalities, leading to increased energy consumption: user tracking,
training of personalised models, and further advertising
technologies that require large-scale infrastructure for storage and data
processing, thereby compounding their environmental footprint.
Additionally, the
sending and rendering of advertisements monopolises resources. In a study
assessing the environmental cost of online advertising, P\"arssinen et al.~\cite{parssinen_environmental_2018} describe the mechanism: when a user visits a web page, it often initiates thousands of connections to data centres on the user’s behalf. Some of these data centres keep the connections open to deliver advertisements as long as the user remains on the page, thereby monopolising network, data centre, and user device resources. Pearce et al.~\cite{pearce2020energy} quantify this waste for user devices in their initiative to estimate the energy savings achievable through the use of ad-blockers, showing that page load times can decrease by up to 28\% with ad-blockers. This represents potential energy savings of 13.5 billion kWh per year for internet users worldwide.

Thus, business models centred on data collection raise not only privacy and
security concerns but also exacerbate environmental costs. Validating the
relationship between business models and energy consumption can guide the
design of more secure, privacy friendly and also more sustainable online
services. This paper compares the environmental burdens of three email
service providers: Microsoft Outlook and Gmail, both of which rely heavily
on advertising and tracking technologies, and Proton Mail, which operates
on a privacy-preserving and end-to-end secure freemium model without the use of advertisements or
trackers.

\section{Methodology}

We propose an approach that builds upon, and extends the work of Pesari et
al.~\cite{pesari2023client}
who examine
the power usage of a computer visiting global news websites when using ad-blockers.
Our approach is designed to assess the energy consumption of a web service by breaking
it down into distinct, measurable functional units. Our framework aims to (i) facilitate the
comparison of similar services based on their energy consumption and (ii)
highlight the connection between data collection, processing, and the
associated environmental costs. We identify three key components in the
assessment: (1) user side, (2) network and server side, and (3) embodied
and end-of-life (EoL) emissions.

\subsection{User-Side Assessment}

To evaluate the emissions released by a user device when accessing
the service, and to identify the attributable share for tracking and
advertising, our framework involves six steps:

\begin{enumerate}
  \item Select Comparable Services: The framework is designed to compare
        different implementations of online services. Different implementations may
        reveal differences in consumption profiles, which could be attributed to
        various factors distinguishing the implementations. However, in our case,
        the aim is not to identify the greener solution, rather to assess the impact
        of advertising mechanisms on device energy consumption. Hence we run the same
        service multiple times, both with and without the ad-blocker activated.
        Additionally we configure the browser to allow the blocking of cookies
        and the restriction of certain website tracking capabilities in order to
        account for the impact of tracking mechanisms.
  \item Define Functional Units: We identify the common scenarios that
        users typically engage in when using the service such as logging into the
        account, reading three emails, replying to, and logging out.
  \item Automate the identified scenarios for repeated execution without
        manual intervention.
  \item Monitor the Scenarios: This step involves executing the scripts
        developed in the previous step and monitoring (i) machine power or energy
        consumption (in Watts or Joules) and (ii) data traffic on the network (in
        Bytes). 
        This is achieved using metric
        tool APIs relying on modern computers' embedded sensors to track their components power consumption.
  \item Convert Energy to Greenhouse Gas (GHG) Emissions: This is typically done using an
        average grid energy mix, which translates to the environmental cost of
        electricity production (expressed in kgCO2-equivalent per kWh).
  \item Assess the User Side Use-Phase Emissions: Perform tasks 1 to 6 for both services and compute the differences in their GHG emissions.
        Additionally, compute the differences in the network data traffic they
        generated.
\end{enumerate}

\subsection{Network and Server-Side Assessment} \label{network-server}


Estimating server-side emissions is challenging due to the lack of
transparency from companies regarding environmental impacts and data
processing and storage. This is highlighted, for example, by the European Data
Protection Supervisor (EDPS) who found that the European Commission’s use
of Microsoft 365 does not comply with EU data protection regulations due to
unclarity regarding the collection, purpose, and processing location of
personal data~\cite{EuropeanCommissionsUse2024}.
Therefore, previous studies assessing the environmental impact of the ICT sector~\cite{malmodin2024ict,freitag_real_2021} often rely on global market and industry statistics for estimates. This leads to significant uncertainty across
different studies. In particular, the impact of training and maintaining advertising models is assumed to have
high impacts that are hard to assess: training AI models is not only known to be a computationally intensive task but also often requires extra hardware accumulating thereby more embodied emissions. As underlined in a review on different estimates of ICT climate impact~\cite{freitag_real_2021}, the rising trend in AI and data science can be regarded as a threat to the future of the ICT climate impact as it drives a steep growth in data storage and processing. Additionally, models are eventually used to target ads more effectively, leveraging the relationship, studied in ~\cite{frick_online_2021}, between perceived consumption-promoting online content and consumption levels. Both kind of impact, i.e., direct and indirect, make it difficult to derive precise estimates of server-side energy consumption for any specific service.
Nevertheless, we include this step to emphasise the importance of
recognising and maintaining awareness of the substantial energy consumption
and associated GHG emissions linked to data processing:

\begin{enumerate}
  \item Collect Data from ICT Sector: We rely on the estimations provided
        in~\cite{malmodin2024ict}, as it contains the most recent data available.
        Specifically, we use the study’s estimates of network and server-side
        emissions per GB of data handled.
  \item Estimate Network and Server-Side Emissions: Multiply the difference
        in data traffic (calculated in step 6) by the respective ratios identified
        in the previous step to estimate the share of attributable network and server emissions.
\end{enumerate}

\subsection{Embodied and End-of-Life Assessment}

Similarly to the previous evaluation, this step involves gathering data
from prior studies that assess the ICT sector~\cite{malmodin2024ict}.
Specifically, we need the ratios of the estimated embodied and End-of-Life
(EoL) emissions for the network and server sides per GB of data handled.

\section{Experimental Results}

We present a first application of our framework to estimate and
compare the environmental footprint of three major email providers:
Microsoft Outlook~\cite{outlook},
Gmail~\cite{gmail}, and Proton
Mail~\cite{proton}. We aim to compare these
services and assess the impact of browsing with and without an ad-blocker,
thus currently focusing on the user side.

We first created different email accounts with each provider. Six
fundamental functional units were defined: \texttt{Idle}, \texttt{Login}, \texttt{Logout}, \texttt{No attachment}, \texttt{Attachment}, \texttt{Reply}, and \texttt{Delete}. The idea is to estimate
consumption for more complex functional units by aggregating the
performance of basic ones.

We then developed a simulation tool
using Selenium~\cite{selenium} in order to automate
scenarios such as mouse clicks, user interactions, logging in, etc. This tool
is available~\cite{artifact-jason}.
For our evaluation we use the Green Metric Tool (GMT)~\cite{greencoding} which is specifically designed to monitor the power consumption of software. To achieve this it uses containers which are small, lightweight virtual environments that contain the necessary dependencies for programs to run. By isolating the program in a container, the GMT creates a controlled environment that accurately evaluate the specific energy consumption and data transfer attributable to the container. As an indicator, the tool automatically converts the data traffic exchanged by the program on the network into the corresponding shares of the network's and server's emissions, which corresponds to the network and server-side assessment[\ref{network-server}] in our
framework. Finally, it can also accept relevant parameters describing the
machine on which it runs to estimate the monitored program’s share of
embodied emissions. This allows an assessment of the user side embodied
emissions. We conducted five runs for each functional unit, collecting data
on CPU energy, memory energy, machine energy, and network I/O. While no
significant differences were found, suggesting that ad-blockers did not
notably affect energy consumption, comparison among providers showed some
differences, with Gmail being slightly more energy-efficient and Proton
consuming more energy, particularly in memory usage. We then provide
estimates for emissions based on energy consumption. An objective is to extend these measurements and contribute to the Energy ID project~\cite{greencoding:energyid} consisting in developing impact-based score cards for different software services.
%

\subsection{Limitations and Future Directions}

The primary limitation of this work, as previously mentioned, is the lack of data transparency from companies handling user information. While monitoring network exchanges can provide an estimate of the data collected by these platforms, there is no reliable way to assess the environmental impact of the data processing happening behind the scenes—simply because we don’t know what is that processing. Enhancing our framework could involve listing the various data storage and processing technologies we are aware of, along with approximate estimates of their carbon cost per gigabyte. This might help us gauge how far a given provider's practices deviate from our current estimates. However, such insights would remain indicative at best and could not serve as a basis for direct comparison. Our view is that, without regulations requiring stakeholders in this industry to disclose
much more detailed information about their activities, it will remain challenging to achieve a
comprehensive and accurate estimate.

In terms of network consumption, we believe our estimates are closer to reality. However, as Freitag et al. discuss in their comparative analysis of ICT climate impact estimates~\cite{freitag_real_2021}, there is ongoing debate in the field regarding this issue. We recognise that network consumption is not elastic, which challenges the validity of our emission allocation. Networks are designed to handle peak activity, consuming a constant amount of energy regardless of data traffic levels. It is only when growing data demand requires infrastructure expansion that emissions increase. Nevertheless, since higher data transfer eventually results in higher emissions, we maintain that allocating responsibility for network emissions is necessary, with the volume of data exchanged serving as a suitable criterion.

On the user side, for which we have the most reliable estimates, several
limitations need to be addressed. First, our tool currently does not
account for the impact of the ad-blocker itself. This limitation could
potentially be addressed by incorporating an external ad-blocker such as
Pi-hole~\cite{OverviewPiholePihole} which blocks ads at the network level
by supplying non-routable DNS entries for ad servers. Additionally, to ensure a fair comparison among providers, functional units have been standardised to operate for the same duration across different providers using wait statements. While this approach helps maintain consistency, it reduces the click rate and may hide the time savings achieved by the ad-blocker, as highlighted by Pearce et al.~\cite{pearce2020energy}. To cross-validate our findings, we could re-run the tests without these wait statements and place greater focus on the time savings aspect.
Another reason for not observing significant differences between runs with
and without ad-blockers could be that web-mail services typically display
fewer ads, relying instead on data collection and reselling as their
business model. Ad-blockers are less effective against such mechanisms as
most processing happens server-side. This could explain the lack of
variation in our current results. We are also critically aware of
projections claiming that between 50\% and 90\% of all email traffic is
spam~\cite{statistaspam2024,fu2014detecting}, the management
of which might have a measurable impact on sustainability metrics of a mail
service. We leave assessment and evaluation of these aspects for future
work.

\section{Conclusions}

Our work offers insights into the environmental impact of online
services. We proposed an approach and framework for the impact
assessment of software systems and partially apply this to evaluate
services such as web mailers. Our approach not
only facilitates comparison between similar services based on energy
consumption, but also highlights the connection between data collection,
processing, and their environmental costs. Ongoing research focuses on
expanding sample sizes, incorporating more diverse scenarios for functional
units, and including server-side data for greater accuracy, and to make the
results more accessible to decision makers.

\begin{acks}
  We gratefully acknowledge the Brussels-Capital Region - Innoviris for financial support under grant numbers 2024-RPF-2 and 2024-RPF-4, and the
CyberExcellence programme of the Walloon Region, Belgium (grant
2110186).
\hyphenation{Innoviris}

\end{acks}
\balance

\bibliographystyle{ACM-Reference-Format}
\bibliography{sustainable-ict.bib,tools.bib}
\balance

\end{document}